# Electric-Field-Controlled Altermagnetic Transition for Neuromorphic Computing

Zhiyuan Duan[1,4], Peixin Qin[1,4,*], Chengyan Zhong[2], Shaoxuan Zhang[3], Li Liu[1,4], Guojian Zhao[1,4], Xiaoning Wang[5], Hongyu Chen[1,4], Ziang Meng[1,4], Jingyu Li[1,4], Sixu Jiang[1,4], Xiaoyang Tan[1,4], Qiong Wu[3,*], Yu Liu[2,*], Zhiqi Liu[1,4,*]

[1]School of Materials Science and Engineering, Beihang University; Beijing 100191, China

[2]College of Integrated Circuit Science and Engineering, Nanjing University of Posts and Telecommunications; Nanjing 210023, China

[3]School of Materials Science and Engineering, Beijing University of Technology; Beijing 100124, China

[4]State Key Laboratory of Tropic Ocean Engineering Materials and Materials Evaluation, Beihang University; Beijing 100191, China

[5]The Analysis & Testing Center, Beihang University; Beijing, 100191, China

*Emails: qinpeixin@buaa.edu.cn; wuqiong0506@bjut.edu.cn; y-liu-17@tsinghua.org.cn; zhiqi@buaa.edu.cn

**ABSTRACT**

Altermagnets represent a novel magnetic phase with transformative potential for ultrafast spintronics, yet efficient control of their magnetic states remains challenging. We demonstrate an ultra-low-power electric-field control of altermagnetism in MnTe through strain-mediated coupling in MnTe/PMN-PT heterostructures with negligible Joule heating. Application of +6 kV/cm electric fields induces piezoelectric strain in PMN-PT, modulating the Néel temperature from 310 to 328 K. As a result, around the magnetic phase transition, the altermagnetic spin splitting of MnTe is reversibly switched "on" and "off" by the electric fields. Meanwhile, the piezoelectric strain generates lattice distortions and magnetic structure changes in MnTe, enabling up to 9.7% resistance modulation around the magnetic phase transition temperature. Leveraging this effect, we implement programmable resistance states in a Hopfield neuromorphic network, achieving 100% pattern recognition accuracy at ≤40% noise levels. This approach establishes the electric-field control as a low-power strategy for altermagnetic manipulation while demonstrating the viability of altermagnetic materials for energy-efficient neuromorphic computing beyond conventional charge-based architectures.

**Introduction**

Altermagnetism has rapidly emerged as one of the most prominent research frontiers in condensed matter physics over the past five years.[1] It was found that altermagnets could constitute a distinct magnetic classification beyond traditional ferromagnets and antiferromagnets. Furthermore, altermagnets, characterized by a collinear and compensated spin structure, are distinct from the emerging class of noncollinear antiferromagnets, which feature noncollinear spin arrangements. While sharing the collinear antiferromagnetic-like arrangement of magnetic moments, altermagnets are marked by a peculiar symmetry breaking in both real and momentum space, arising from specific crystal structures. This unique symmetry enables ferromagnetic-like phenomena such as significant spin splitting and corresponding magneto-electronic effects.[2,3] Theoretical calculations predict altermagnetism across multiple material classes, including metals, semiconductors, and insulators. To date, several candidate materials, *e.g.*, $RuO_2$, CrSb, MnTe, $Mn_5Si_3$, $KV_2Se_2O$ and $Rb_{1-\delta}V_2Te_2O$, have been experimentally validated as altermagnets through intensive research.[4–9] In addition, the observation of anomalous Hall effect,[4,10–13] spin-splitting magnetoresistance,[14] spin-splitting torque[15–17] and tunneling magnetoresistance effect[18] positions altermagnets as promising systems for next-generation spintronics.

More importantly, compared to ferromagnets, altermagnets offer compelling advantages for spintronic applications due to their antiferromagnetic-like order: ultrafast spin dynamics (up to ~THz), negligible stray fields, and robustness against interference.[19,20] However, effectively modulating their magnetic states remains a significant challenge for both fundamental studies and device implementation at current stage. Owing to antiferromagnetic coupling, altermagnets exhibit a very high critical magnetic switching field, comparable to that of conventional collinear antiferromagnets, with predicted values reaching several or even tens of tesla.[19,20] For instance, in altermagnetic RuO$_2$ thin films,[4,13] the anomalous Hall effect has been observed under an external magnetic field as high as 50 T. As a result, it is extremely challenging to manipulate the spin state of altermagnets via conventional magnetic field approaches. Therefore, electrical control of the magnetic order in altermagnets represents a crucial prerequisite for device integration. Recently, spin-orbit torques induced by applying an electric current have been successfully employed to manipulate the Néel vector in altermagnets such as Mn$_5$Si$_3$[13] and CrSb[21]. In these studies, the current density typically reaches values on the order of $10^7$ mA/cm$^2$, which could lead to significant energy dissipation and substantial Joule heating. Notably, unlike conventional control methods such as magnetic fields and electrical currents, electric field can offer an energy-efficient approach with negligible heat dissipation,[20,22] particularly suited to antiferromagnetic and altermagnetic materials requiring high critical switching current densities. For instance, electric-field-induced piezoelectric strain has enabled efficient manipulation of both collinear antiferromagnets (MnPt and MnIr) and noncollinear antiferromagnet (Mn$_3$Pt, Mn$_3$Sn, Mn$_3$Ga and Mn$_3$Ge), producing significant functional outputs.[23–28] More importantly, the multilevel resistance states achieved through electric-field control are well suited for directly representing analog synaptic weights in neuromorphic computing.

This makes MnTe—an established altermagnetic candidate with experimentally confirmed spin splitting[6,29] and anomalous Hall signatures[12]—a particularly compelling testbed for novel control strategies. As a conventional metal chalcogenide with a high melting point of 1165 °C,[30] it has been primarily investigated for thermoelectric applications due to its tellurium content.[31] At room temperature, it crystallizes in the hexagonal α-MnTe phase (space group *P*6$_3$/*mmc*; Strukturbericht designation B8$_1$, Figure 1b). Historically classified as a collinear antiferromagnet and *p*-type semiconductor,[6,32] its low conductivity hinders efficient manipulation via conventional charge- or spin-current methods. Besides, its metal-like transport manifests at low temperatures alongside a Néel transition at approximately 310 K.[6,33]

Recently, theoretical studies propose that strain and pressure can effectively modulate altermagnets, offering a viable approach for controlling MnTe systems.[34–36] Piezoelectric strain could offer a practical pathway for magnetic structure control unlike conventional static-pressure approaches demanding tens of kbar.[37–40] Motivated by these points, in this work, we synthesized high-quality altermagnetic MnTe single crystals and fabricated electric-field-controlled heterostructures integrating MnTe flakes with ferroelectric $0.7PbMg_{1/3}Nb_{2/3}O_3$–$0.3PbTiO_3$ (PMN–PT). The application of an electric field of +6 kV/cm enables an 18 K modulation of the Néel temperature, shifting from 310 to 328 K. We further demonstrated neuromorphic computing functionality through Hopfield associative memory networks implemented via programmable MnTe resistance states achieving impressive accuracy under both lower and higher noise levels.

**Results and discussion**

High-purity MnTe single crystals were synthesized via a modified Bridgman technique (Figure 1a, (see the Supporting Information for details). Figure 1c shows the optical image of the synthesized MnTe single crystal. Powder X-ray diffraction (XRD) analysis (Figure 1d) reveals exclusively α-MnTe phase reflections (space group $P6_3/mmc$), attesting to phase purity. Rietveld refinement implemented in GSAS-II yields lattice parameters $a$ = 4.1482 Å and $c$ = 6.7125 Å, demonstrating striking consistency with established crystallographic data.[41,42]

Atomic-resolution scanning transmission electron microscopy (STEM) elucidates the hexagonal lattice symmetry (Figure 2a-c) in [100] crystallographic direction. Complementary electron energy loss spectroscopy (EELS) mapping (Figure 2d) confirms stoichiometric Mn: Te distribution.

Moreover, high-angle annular dark-field (HAADF) imaging (Figure 2e-h) resolves distinct Mn and Te sublattices, revealing a 1:1 periodicity ratio between Mn and Te atomic planes - a hallmark of the $\alpha$-MnTe structure. This comprehensive structural characterization unambiguously verifies the synthesis of phase-pure, single-crystalline MnTe.

Previous studies established that $\alpha$-MnTe undergoes an altermagnetic-to-paramagnetic transition at ~310 K.[6,33,43] We corroborated this behavior through both magnetization and resistivity measurements. The magnetic-field dependent magnetization measurements were performed along the [100] crystallographic axis of MnTe (Figure S1). The observed linear dependence and the small magnitude of the magnetization are characteristic of a compensated magnetic system with negligible net moments, thereby confirming the altermagnetic nature of MnTe. As shown in Figure 3a, temperature-dependent magnetization measurements were performed under a 0.1 T magnetic field applied along the same direction. The magnetization curve exhibits an upward trend followed by a downward trend as temperature rises, with the turning point starting at ~310 K (indicated by arrow and gradient fill), characteristic of the Néel transition. As plotted in Figure 3b, temperature-dependent resistance measurement reveals analogous critical behavior near Néel temperature (~310 K), though with persistent increase beyond the transition temperature accompanied by slope modulation. The magnitude of resistivity is about 0.6 $\Omega\cdot$cm at room temperature, showing the poor conductivity of MnTe and consistent with the traditional semiconductor nature.

To investigate electric-field control of MnTe and its spintronic applications, we engineered a prototype device integrating MnTe flake with $0.7PbMg_{1/3}Nb_{2/3}O_3$–$0.3PbTiO_3$ (PMN-PT) substrate. The device and measurement geometry are illustrated in Figure 4a. When an electric field is applied perpendicularly to the device, it induces piezoelectric strain in the PMN-PT substrate. This strain propagates to the overlying MnTe flake, as illustrated in Figure S2a, generating controlled deformation and microscopic lattice distortions/magnetic structure changes that modulate its magnetic and electrical properties. As shown in Figure 4b, when an electric field of +6 kV/cm was applied, the temperature-dependent resistance reaches its maximum value at 328 K, which indicates that the Néel temperature lift up 18 K compared to the voltage-free circumstance. A possible explanation is that the strain-induced strengthening of the dominant antiferromagnetic exchange interaction increases the energy required to destroy the magnetic order, thereby stabilizing the altermagnetic phase. Therefore, the application of a +6 kV/cm vertical electric field can enable effective switching between altermagnetic and non-altermagnetic states in MnTe at intermediate temperatures (310–328 K), corresponding to the on-off switching of the altermagnetic spin splitting around its magnetic phase transition temperature. This result achieves comparable Néel temperature modulation in MnTe to prior studies employing static pressure up to tens of kbar.[37–40] Besides, it can be anticipated that the effectiveness of the electric-field control could be further enhanced when the sample is in a single-domain state. Regarding the lower temperature, the resistance of MnTe exhibited negligible change. On the one hand, it could because that the coercive field of PMN-PT for polarization switching increases and consequently transition is suppressed at low temperatures. On the other hand, the most dramatic resistance changes typically occur near phase transitions *i.e.*, Néel temperature, resulting in negligible resistance variation away from the transition point.[25,44]

Figure 4c displays the temperature-dependent resistance difference ($\Delta R$) between zero-field and +6 kV/cm electric-field states of MnTe. This differential measurement quantifies strain-induced modulation in the MnTe flake across the thermal range, demonstrating the control capability of our device. Moreover, at 328 K, the $\Delta R$ reached its maximum value of ~1.0 Ω, and the changing ratio is approximately 9%, which is comparable to the previous antiferromagnets with phase transition such as FeRh from antiferromagnet to ferromagnet,[44] and Mn$_3$Pt from noncollinear antiferromagnet to collinear antiferromagnet.[23] When the temperature rises above 328 K, the $\Delta R$ declined, which further indicated that the electric-field modulation of resistance changes most abruptly around the phase-transition point. It is noteworthy that experimental studies on the electric-field control of altermagnets are currently scarce. Given the vast number of altermagnet candidates and their rich diversity of magnetoelectric responses, electric-field manipulation could be pivotal for further deepening understanding of intrinsic mechanisms of altermagnets and advancing their practical applications.

Furthermore, the strain variation and longitudinal resistance at different temperatures (300, 310, 315, 320, and 328 K) around the phase change temperature have been collected. Figure 5a exhibits two distinct peaks in gate current, unambiguously confirming polarization switching behavior in the ferroelectric PMN-PT. The spike-shape change originates from the coercive electric field near the flip of the ferroelectric domains. It should be noted that the leakage current remains on the order of nA, resulting in negligible Joule heating. This minimal energy dissipation highlights the eco-efficient nature of electric-field control, enabling energy-efficient, electrical-based programming of the synaptic devices. Across all measured temperatures, the electric-field dependent four-probe resistance (Figure 5b-f) exhibits characteristic butterfly hysteresis loops, featuring opposing spikes at about ±1 kV/cm that align precisely with leakage current maxima in Figure 5a. The obtained butterfly hysteresis loops indicate the multilevel resistance states, which are ideal for directly representing analog synaptic weights in neuromorphic networks. As shown in Figure 5b, the strain induced by an electric field of +6 kV/cm is about 0.053% at 300 K, leading to an electroresistance of ~6.5%. The change of longitudinal resistance can originate from a reorientation of the Néel vector in MnTe, mediated by the applied piezoelectric strain. At 328 K, the piezoelectric strain is 0.069% and the resistance change ratio raise up to 9.7% (Figure 5f). These two intrinsic characteristics, multilevel analog states and low operating power, together establish a solid foundation for the implementation of neuromorphic computing systems that are both energy-efficient and highly accurate.

Moreover, two distinct remnant strain states are clearly resolved at zero electric field, indicating that the modulation of strain in PMN-PT can result in a nonvolatile resistivity change. The nonvolatile piezoelectric strain in PMN-PT originates from 109° ferroelectric domain switching, characterized by a 90° rotation of the spontaneous polarization component within the (001) plane. As shown in Figure S2b, the spontaneous ferroelectric polarization easy axis is aligned along the <111> family of directions, resulting in eight equivalent polarization orientations. Using the [111] direction as an initial example, when an external electric field is applied along the [001] crystallographic direction, the polarization can be rotated to other directions by 71°, 109° or 180°, which in our schematic stands for [11$\bar{1}$], [$\bar{1}$1$\bar{1}$], and [$\bar{1}\bar{1}\bar{1}$], respectively. Given that the PMN-PT crystal PMN-PT crystal is [001]-oriented and the device is fabricated on the (001) plane, the polarization directions must be projected onto this plane. The original polarization direction is projected to [110]. When the polarization direction changes, for 109° rotation, the switch can cause the in-plane polarization change from [110] to [$\bar{1}$10] in (001) plane, thereby causing a 90° change in the polarization direction.[45] In contrast, 71° and 180° rotations do not cause change in the polarization direction. It is this specific 109° switching that underlies the nonvolatile resistance behavior observed in our devices.[46,47]

To demonstrate the practical applicability of the electric-field-controlled altermagnetic transition for neuromorphic computing, we implemented a Hopfield associative memory network using the extracted resistance values from MnTe devices as synaptic weights (gray circles in Figure 5f).

Figure 6a illustrates the network architecture, where input images are flattened into binary vectors that feed into a fully connected network with visible and hidden layers interconnected through weight matrix *W*. The network operates as an associative memory system, capable of reconstructing corrupted or incomplete input patterns by converging to stored memory states. The network implementation employs $N = 16384$ binary neurons with symmetric weight connections trained via Hebbian learning rules. To emulate practical hardware constraints, weight quantization mechanisms are applied to discretize continuous synaptic values onto finite precision levels. Energy consumption analysis estimates approximately 1 fJ per synaptic read operation, yielding ~16 pJ per full network iteration for our implementation. Comprehensive details of the network architecture, training algorithms, quantization procedures, and energy calculations are provided in Supplementary Note of the Supporting Information. Figure 6b demonstrates the network's denoising capability across multiple test images, comparing the original clean images, their corrupted versions with added noise, and the reconstructed outputs from both the MnTe device-based implementation and software simulation. The results shown correspond to 40% noise corruption applied to three representative test patterns: Cameraman, Astronaut, and Text, where the substantial noise corruption renders the input images nearly unrecognizable. The device-based reconstruction shows remarkable fidelity to the software implementation, successfully recovering recognizable patterns from heavily corrupted inputs, which validates the feasibility of using altermagnetic resistance states as analog weights in neuromorphic systems.

The quantitative performance analysis reveals excellent agreement between hardware and software implementations across varying noise levels. Figures S3-S5 provide detailed visualization of the reconstruction process across different noise intensities (10%-80%), showing the progressive degradation of input quality and the corresponding device-based outputs, which maintain structural integrity even under substantial corruption. Figure 6c presents the normalized weight distribution extracted from the MnTe devices, displaying a continuous range of resistance states mapped to synaptic strengths between -1 and 1. The accuracy comparison in Figure 6d shows that the device-based Hopfield network achieves performance nearly identical to software simulation, with reconstruction accuracies reaching 100 % for low noise levels (10%-40%) and maintaining over 40% accuracy even at 50% noise corruption for both implementations. To comprehensively characterize the performance trade-offs, we conducted systematic Pareto analysis examining energy efficiency, memory capacity, and retrieval accuracy. Figure S6 illustrates how recall accuracy degrades as the number of stored patterns $P$ increases across different noise levels (0%-40%). The results reveal a clear capacity limit: when the storage ratio $α = P/N$ approaches approximately 0.138 (corresponding to $P ≈ 2,300$ for our $N = 16,384$ network), accuracy begins to decline rapidly regardless of noise level, consistent with theoretical predictions for Hopfield networks. Figure S7 presents the energy-accuracy relationship during inference under 30% noise corruption, demonstrating that the system achieves 95% accuracy with merely ~0.1 nJ energy consumption and reaches near-perfect reconstruction (>99%) at ~0.3 nJ. This steep convergence profile establishes sub-nanojoule operation as a practical regime for the MnTe-based implementation, achieving competitive energy efficiency while maintaining robust associative memory functionality. This remarkable consistency between hardware and software performance, combined with the device's ability to provide stable, analog resistance states through electric-field control, establishes MnTe-based devices as promising candidates for energy-efficient neuromorphic computing applications.

**Conclusion**

This work demonstrates deterministic electric-field control of altermagnetism through strain-engineered MnTe/PMN-PT heterostructures, achieving an 18 K enhancement of the Néel temperature (from 310 to 328 K) at an electric field of +6 kV/cm. The resistance change in MnTe flake originates from strain induced by ferroelectric domain switching in PMN-PT and 9.7% resistance modulation has been achieved near the phase transition point. We further validate neuromorphic functionality through Hopfield networks implementing analog resistance states, achieving 100% pattern recognition accuracy below 40% noise levels. These breakthroughs resolve the persistent challenge of altermagnetic state manipulation while establishing dual-functionality for ultrafast spintronics and energy-efficient neuromorphic computing. The piezoelectric strain modulation mechanism provides a universal platform for electric-field control of symmetry-broken quantum materials. Moreover, beyond $\alpha$-MnTe in this work, future development of practical altermagnetic devices could prioritize candidate materials that exhibit a Néel temperature near or above room temperature, strong electric-field tunability, and excellent ambient stability. Specifically, a temperature near room temperature enables remarkable signal output, while a higher temperature ensures broader operational feasibility. The electric-field response, a critical feature for device application, demands thorough investigation to meet specific performance requirements. Based on these considerations, materials such as $Rb_{1-\delta}V_2Te_2O$, $KV_2Se_2O$, $RuO_2$, and CrSb emerge as promising candidates for further exploration.[4–10, 48–50] The FeS system,[51] whose properties can be broadly tuned through composition engineering, also constitutes a compelling target for future studies.

T.; Arita, R.; Seki, S. Spontaneous Hall Effect Induced by Collinear Antiferromagnetic Order at Room Temperature. *Nat. Mater.* **2025**, *24*, 63–68.

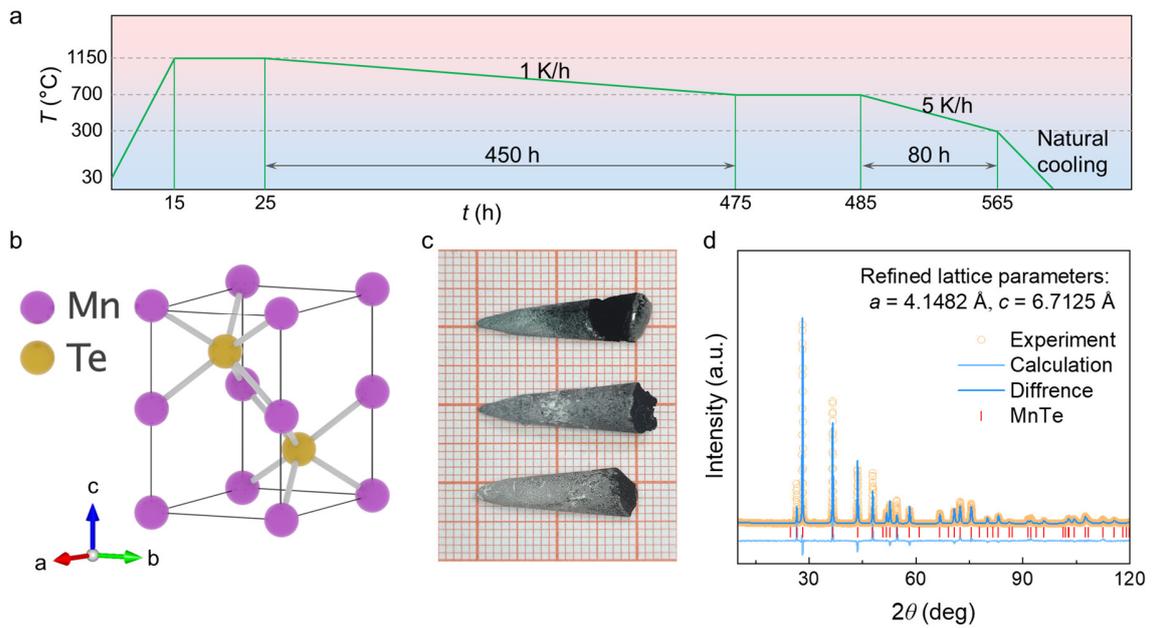

**Figure 1.** (a) Temperature profile for MnTe single crystal synthesis via the Bridgman method. (b) Crystal structure of MnTe. Purple and brown spheres denote Mn and Te atoms, respectively. (c) Optical image of as-grown MnTe single crystals. (d) XRD pattern of ground MnTe with Rietveld refinement results (lattice parameters: $a$ = 4.1482 Å, $c$ = 6.7125 Å).

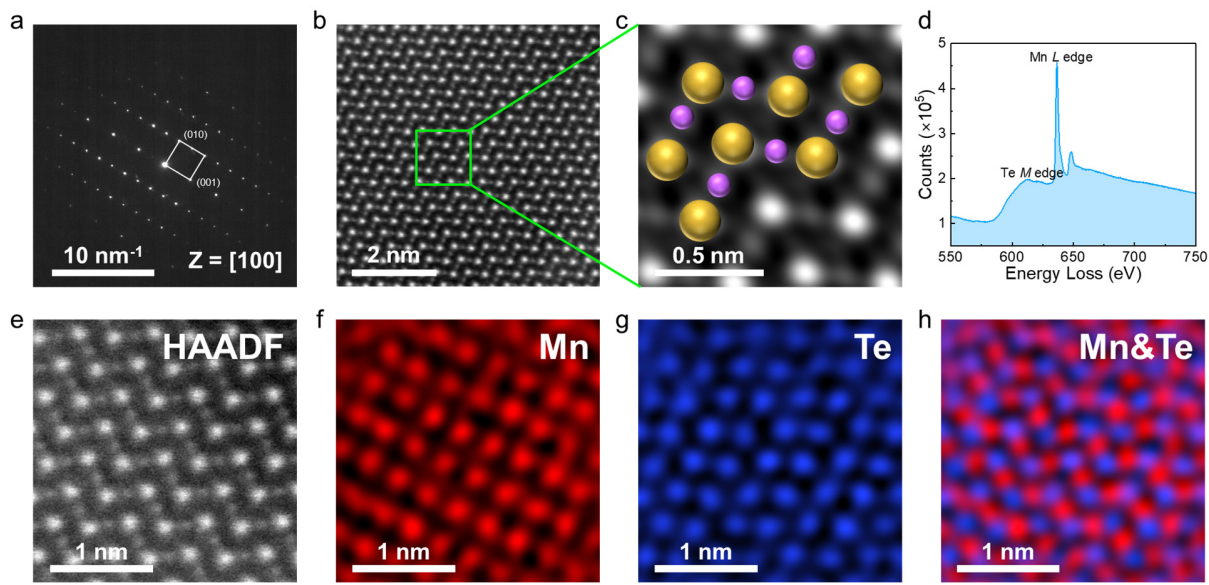

**Figure 2.** (a) SAED pattern of MnTe single crystal along the [100] zone axis. Scale bars: 10 nm⁻¹ (b) STEM-HAADF images of MnTe acquired along the [100] zone axis. Scale bars: 2 nm. (c) Magnified view of the boxed region in (b) overlaid with the atomic structure of MnTe (pink spheres: Mn; brown spheres: Te), Scale bar: 0.5 nm. (d) EELS spectrum showing the Mn L-edge and Te M-edge (energy range: 550–750 eV). (e–h) Elemental mapping of MnTe: (e) HAADF image, (f) Mn map, (g) Te map, (h) Composite of Mn (red) and Te (blue), Scale bars: 1 nm.

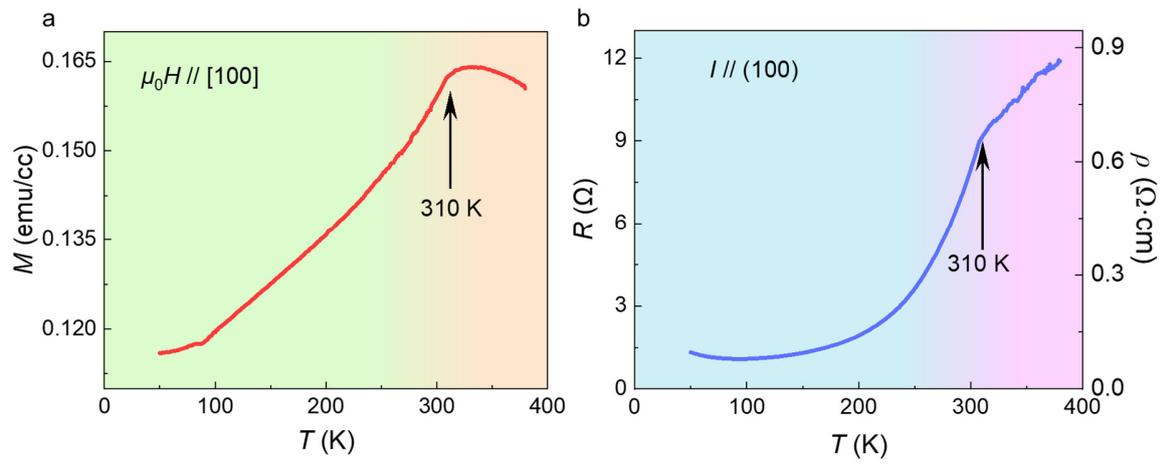

**Figure 3.** (a) Temperature-dependent magnetic moment (*M*) of an MnTe single crystal from 50 to 380 K, showing an altermagnetic transition at 310 K. (b) Temperature-dependent resistance (*R*) of the same crystal from 50 to 380 K, with a corresponding transition at 310 K.

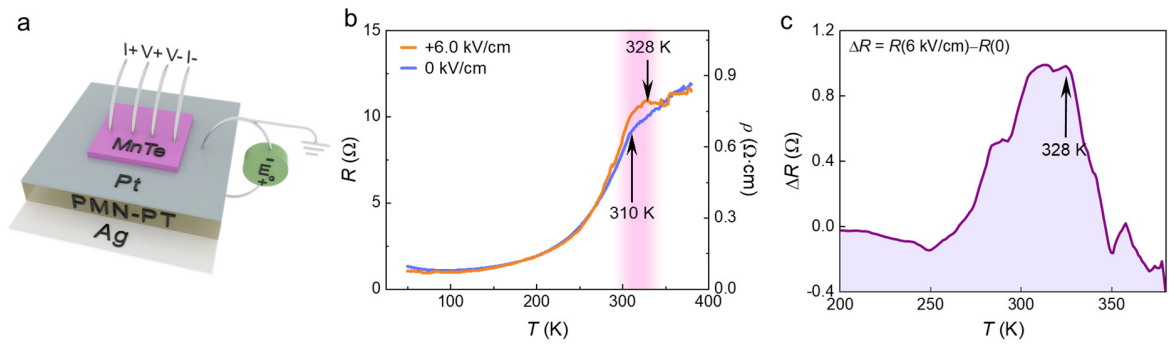

**Figure 4.** (a) Schematic of the MnTe-based electric-field-controlled device. (b) Temperature dependence of resistance under $E$ = 0 kV/cm and +6 kV/cm, with electric-field-tunable transition region highlighted (pink shading). (c) Temperature-dependent resistance difference ($\Delta R$) of MnTe in the critical range 200–328 K.

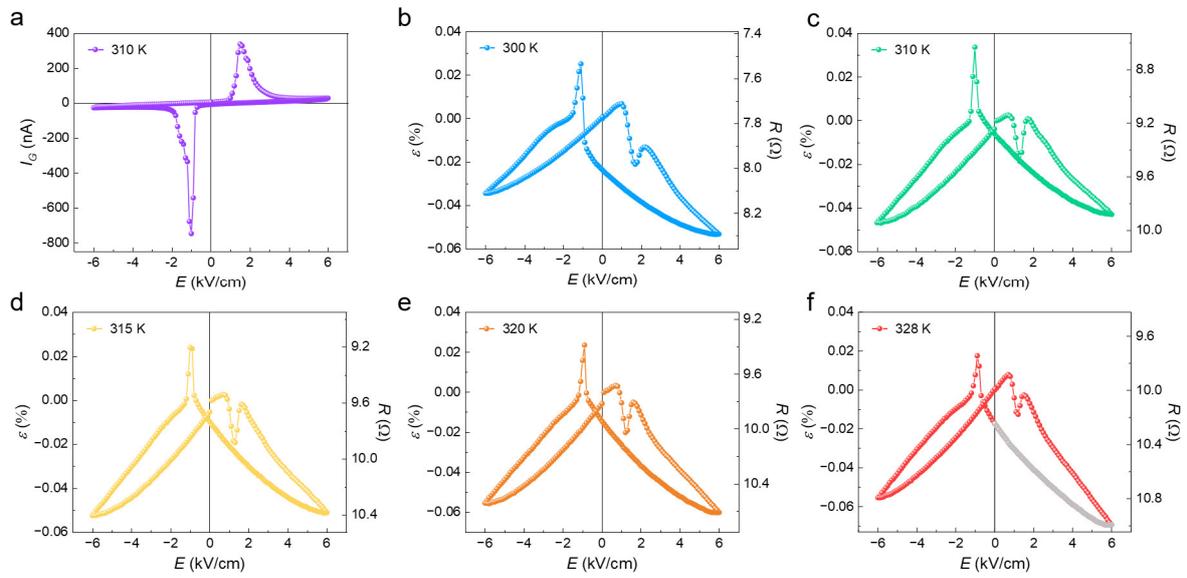

**Figure 5.** (a) Electric field dependence of leakage current under 310 K, (b-f) Electric field dependence of strain coupled with resistance at (b) 300, (c) 310, (d) 315, (e) 320, and (f) 328 K, respectively. The gray balls represent the data points used for the neuromorphic computing.

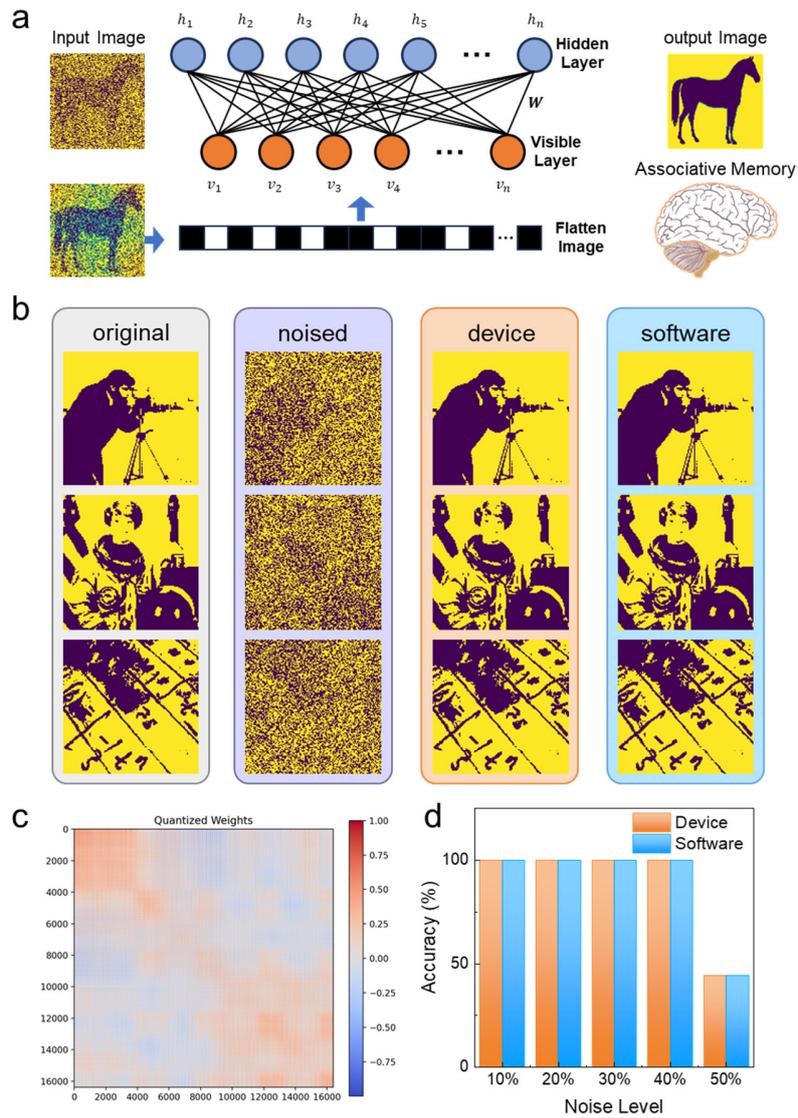

**Figure 6. Hopfield associative memory network implementation using MnTe-based synaptic devices. (a)** Schematic diagram of the Hopfield network architecture showing input image processing through flattened binary vectors, visible layer ($V_1$-$V_n$), hidden layer ($h_1$-$h_n$), and weight matrix W connections for associative memory operation. **(b)** Comparison of image reconstruction performance showing original images, noised input images, device-based reconstruction outputs, and software simulation outputs for three representative test patterns. **(c)** Normalized weight distribution matrix extracted from MnTe device resistance states, with values mapped to synaptic strengths ranging from -1 to 1. **(d)** Reconstruction accuracy comparison between device-based (blue bars) and software-based (orange bars) implementations across different noise levels.

**Electric-Field-Controlled Altermagnetic Transition for Neuromorphic Computing**

*Zhiyuan Duan[1,4], Peixin Qin[1,4],\*, Chengyan Zhong[2], Shaoxuan Zhang[3], Li Liu[1,4], Guojian Zhao[1,4], Xiaoning Wang[5], Hongyu Chen[1,4], Ziang Meng[1,4], Jingyu Li[1,4], Sixu Jiang[1,4], Xiaoyang Tan[1,4], Qiong Wu[3,\*], Yu Liu[2,\*], Zhiqi Liu[1,4,\*]*

[1]School of Materials Science and Engineering, Beihang University; Beijing 100191, China

[2]College of Integrated Circuit Science and Engineering, Nanjing University of Posts and Telecommunications; Nanjing 210023, China

[3]School of Materials Science and Engineering, Beijing University of Technology; Beijing 100124, China

[4]State Key Laboratory of Tropic Ocean Engineering Materials and Materials Evaluation, Beihang University; Beijing 100191, China

[5]The Analysis & Testing Center, Beihang University; Beijing, 100191, China

**Corresponding author**

\*Email: qinpeixin@buaa.edu.cn; wuqiong0506@bjut.edu.cn; y-liu-17@tsinghua.org.cn; zhiqi@buaa.edu.cn

# Experimental Section

## Material growth

MnTe single crystals were synthesized via a simplified Bridgman method. Firstly, two element powder was uniformly mixed with a total mass of 2.5 g, in which the purity of manganese is 99.95% and the purity of Tellurium is 99.99%. The initial composition was controlled at a Mn:Te ratio of 0.49:0.51.[1,2] This Te-rich environment was necessary to promote the formation of the target MnTe phase and minimize $MnTe_2$ impurities, while considering the maximum heating temperature. After that, the mixed powder was loaded in a conical boron nitride crucible and vacuum-sealed in a quartz tube with approximately $10^{-3}$ Pa based pressure. The whole process of sample preparation was repeated identically three times and then the three obtained quartz tubes were placed in an OTF-1200X-S-VT vertical tubular furnace of HF-Kejing. The temperature program of the furnace is set according to the scheme in Figure 1a, which resemble the process we used for $Mn_3Sn$.[3] In the scheme, the furnace was kept at the highest temperature 1150 °C for 10 h to homogenize the MnTe flux. Afterwards, the descending rate was set for 1 K per hour to create the temperature gradient needed for single crystal growth. After sintering and cooling down, the obtained crystal was sliced using a Diamond wire saw and polished with 1000 grit sandpaper.

## Device fabrication

The MnTe single crystal was processed into a smaller and thinner size flake suitable for the electric field modulation via micromachining. A 10-nm-thickness Pt thin film was grown on a (001)-oriented $0.7PbMg_{1/3}Nb_{2/3}O_3$–$0.3PbTiO_3$ (PMN-PT) by a D.C. sputtering system with a base pressure of $10^{-5}$ Pa under room temperature, which serves as the top electrode of our piezoelectric device. And the silver paint is utilized as the bottom electrode. The MnTe flake was transferred on the top of PMN-PT substrate using a cyanoacrylate adhesive.

**Structure characterization**

X-ray diffraction pattern was collected via a Bruker D8 Advance diffractometer with Cu-$K_\alpha$ radiation and the $\lambda$ is 0.15418 nm. Transmission electron microscopy image was captured by a Talos F200X-G2 system operated at 200 kV.

**Magnetic and electrical transport**

Magnetic measurements were carried out via a Quantum Design VersaLab system with a vibrating sample magnetometer utility. A standard four-probe method was utilized for the longitudinal resistance measurements and the electrical contacts were performed by a wire bonding with Al wires (30 μm in diameter). Temperature dependent resistance was measured by the resistance module in VersaLab. For piezoelectric modulation, a Keithley 6221 current source was employed to impress a current on the sample. The gate electric field was supplied by a Keithley 2410 source meter. The voltage of the sample was captured using a Keithley 2182A nanovolt meter.

**Supplementary Note: Hopfield Network Implementation and Energy Analysis**

The Hopfield associative memory network was simulated to evaluate the neuromorphic functionality of MnTe-based synaptic devices.

**Hopfield Network Structure**

The Hopfield network is a fully connected recurrent neural architecture that functions as an energy-based associative memory model. It consists of $N$ binary neurons, each taking a value $s_i \in \{-1, +1\}$. Every pair of distinct neurons is symmetrically connected with a weight $W_{ij} = W_{ji}$, while self-connections are prohibited by setting $W_{ii} = 0$. The overall network state can therefore be represented as a vector:

The implementation followed a fully connected recurrent architecture with $N = 128 \times 128$ binary neurons, each capable of adopting ±1 states. A total of $P = 8$ binary patterns were encoded as reference memories, including representative grayscale images ("Camera", "Astronaut", "Horse", "Coffee", "Coins", "Text", "Checker" and "Brick") binarized at 0.5 threshold.

Each grayscale image was first resized to a uniform resolution of 128×128 pixels to ensure consistent dimensionality across all samples. The pixel intensities were then binarized using the mean threshold method, converting the image into a binary pattern representing bright and dark regions. To match the requirements of Hopfield network representation, the binary values were mapped from {0,1} to {−1, +1}, where +1 denotes an active neuron (white pixel) and −1 denotes an inactive neuron (black pixel). The resulting two-dimensional pattern was subsequently flattened into a one-dimensional vector of length $N = 128^2 = 16,384$, forming the input state of the network. Each of these vectors was then stored as a distinct memory pattern during the weight training phase.

**Noise**

To evaluate the associative recall capability of the Hopfield network under imperfect conditions, controlled noise was introduced into the input patterns before retrieval. Each original binary image pattern $\mathbf{x} \in \{-1, +1\}^N$ was partially corrupted by randomly flipping the sign of individual neuron states according to a predefined corruption probability $p$. Formally, the corrupted input $\mathbf{x}'$ was generated as

$$x'_i = \begin{cases} -x_i, & \text{with probability } p, \\ x_i, & \text{with probability } 1 - p, \end{cases}$$

where $p \in [0.1, 0.8]$ corresponds to a 10–80% noise level. This bit-flip perturbation simulates random pixel inversion in the binary image representation, effectively modeling partial loss or distortion of the stored pattern. The corrupted inputs were then used to test the network's ability to recover the original memories, thereby quantifying its robustness against increasing levels of input degradation.

**Training**

The Hopfield network was trained to store a set of eight binary image patterns, each represented as a vector of neuron states $\mathbf{x}^{(\mu)} \in \{-1, +1\}^N$, where $N = 128^2$. The training process followed the standard Hebbian learning rule, in which the symmetric weight matrix $\mathbf{W}$ was computed as

$$\mathbf{W} = \sum_{\mu=1}^{P} \mathbf{x}^{(\mu)} \mathbf{x}^{(\mu)T}, W_{ii} = 0,$$

with $P = 8$ denoting the total number of stored patterns. This formulation enables each image vector to form an attractor state within the network's energy landscape, allowing the system to recover the original pattern when presented with a corrupted or incomplete version during recall.

**Weight Quantization Mechanism**

To emulate the finite precision of physical hardware implementations such as memristor crossbar arrays or digital synaptic circuits, a weight quantization mechanism is integrated into the Hopfield model. After the weight matrix $W$ is trained, all continuous-valued weights $W_{ij} \in [-1,1]$ are discretized onto a set of predefined quantization levels:

$$\mathcal{L} = \{l_1, l_2, \ldots, l_K\}, l_k \in [-1,1].$$

Each weight is replaced by its nearest quantized level according to the minimum Euclidean distance criterion:

$$W_{ij}^{(q)} = \arg \min_{l_k \in \mathcal{L}} \mid W_{ij} - l_k \mid.$$

This quantization process effectively simulates limited-bit weight storage in neuromorphic hardware, allowing the investigation of precision-induced performance degradation and robustness under discrete weight constraints.

**Energy Estimation**

By utilizing standard Keithley pulse current sourcemeter and nanovoltmeter, the current can be reduced to as low as 1 µA with pulse widths of approximately 100 µs. The energy cost per synaptic read operation was estimated from the following experimental parameters: average read current $I$ = 1 µA, device resistance $R$ = ~10 Ω, and read duration $t$ = 100 µs.

The energy per read operation is given by:

$$E_{read} = I^2 \times R \times t$$

Substituting the values:

$E_{read} = (1 \times 10^{-6} A)^2 \times 10\, \Omega \times 100 \times 10^{-6} s = 1$ fJ per synaptic operation.

For our Hopfield network processing 128×128 binary images, the network contains $N$ = 16384 neurons. During each inference iteration, all neural states must be updated by reading their corresponding MnTe device states. The energy per iteration is:

$E_{iteration} = 1$ fJ $\times 16384 = 16.4$ pJ per iteration.

**Memory Capacity & Accuracy**

To further evaluate the associative performance of the Hopfield network, we conducted a series of simulations analyzing the trade-offs between memory capacity and accuracy. In these simulations, we examined how the average recall accuracy varies as the number of stored patterns $P$ increases and as different levels of corruption are applied to the input states. The results are summarized in Figure S6.

The figure illustrates the average recall accuracy of a classical Hopfield network under different storage loads and noise levels. As the number of stored patterns $P$ increases, the network experiences higher pattern interference, leading to a gradual decline in associative recall performance. Similarly, increasing the input corruption ratio (noise level) reduces the accuracy of retrieved patterns due to the amplified difficulty of converging to the correct memory attractor. Across all noise levels, a clear capacity limit is observed: when the storage ratio $\alpha = P/N$ approaches the theoretical bound of approximately 0.138, the recall accuracy begins to degrade rapidly. This demonstrates the classical trade-off between memory load, input noise, and associative retrieval capability in Hopfield-type attractor networks.

**Energy efficiency & Accuracy**

To quantify the energy-accuracy trade-offs in our MnTe-based implementation, we analyzed the energy consumption during the inference process. Figure S7 shows the relationship between total energy expenditure and achieved accuracy for pattern retrieval under 30% noise corruption.

The curve demonstrates the convergence dynamics of the network: starting from an initial accuracy of approximately 79%, the system rapidly improves to near-perfect reconstruction (>99%) within approximately 0.3 nJ of energy consumption. The steep initial rise indicates efficient energy utilization, with the network achieving 95% accuracy using only ~0.15 nJ. Beyond this point, marginal accuracy improvements require progressively more energy, establishing a practical operating point for energy-constrained applications. Notably, this energy efficiency positions the MnTe-based implementation competitively against existing neuromorphic platforms, while offering additional advantages of non-volatile storage and analog weight tunability.

**Figure S1**

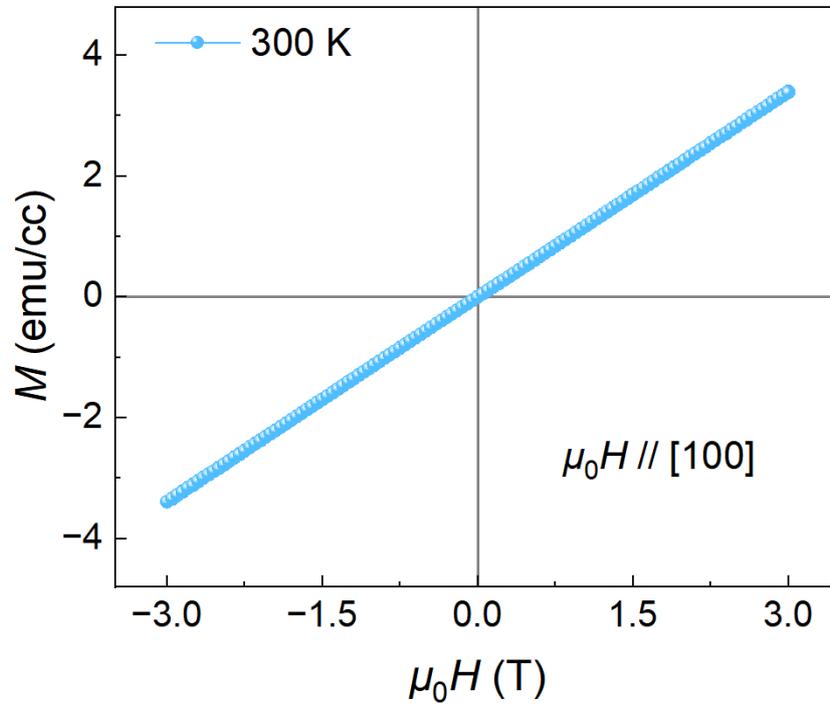

**Figure S1**. The magnetic field-dependent magnetization in MnTe along the [100] crystallographic direction.

**Figure S2**

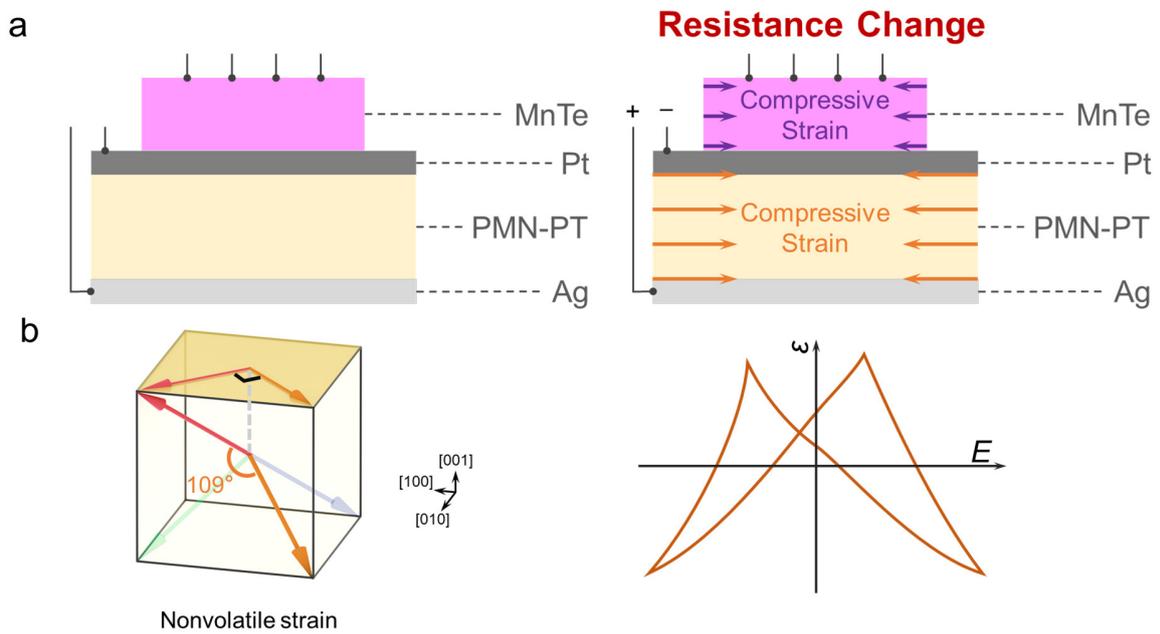

**Figure S2.** (a) Illustration of electric field induced strain effect in the device. (b)Schematic of ferroelectric domain switching processes of PMN-PT.

**Figure S3-S5: Detailed visualization of Hopfield network reconstruction under varying noise conditions.**

**Figure S3**

**Figure S3.** Original and input images with progressive noise corruption levels from 10% to 80%.

**Figure S4**

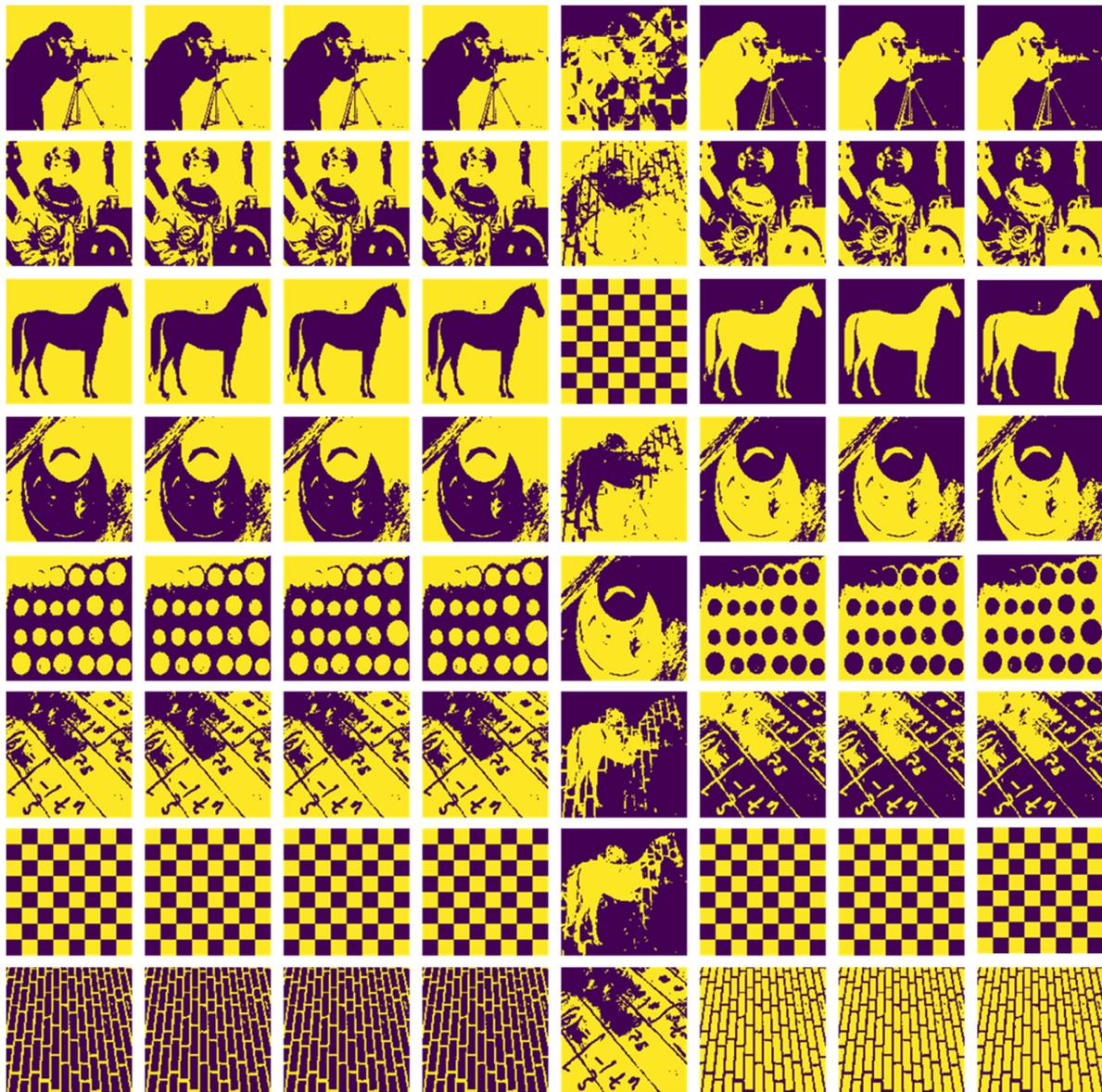

**Figure S4.** Device-based reconstruction outputs with progressive noise corruption levels from 10% to 80%, demonstrating the network's denoising capability and structural pattern preservation across different corruption intensities.

**Figure S5**

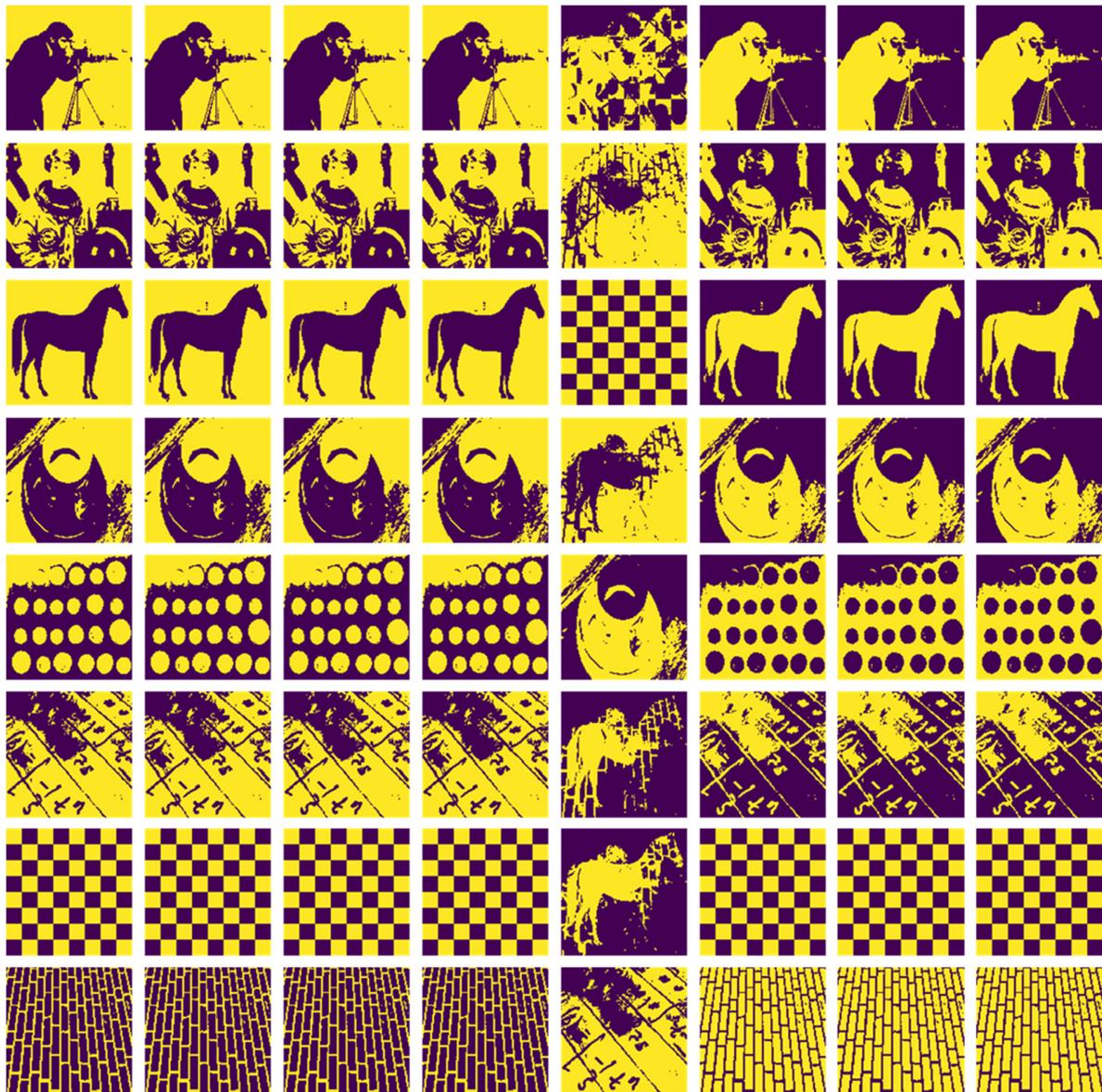

**Figure S5.** Software-based reconstruction outputs with progressive noise corruption levels from 10% to 80%, demonstrating the network's denoising capability and structural pattern preservation across different corruption intensities.

**Figure S6**

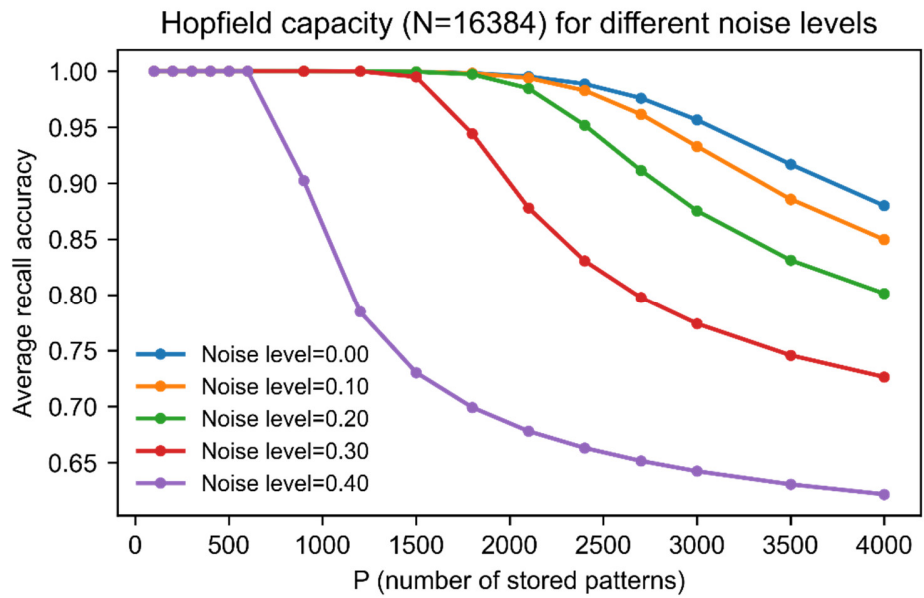

**Figure S6.** Average recall accuracy of a Hopfield network ($N = 16384$) under different memory loads and noise levels.

**Figure S7**

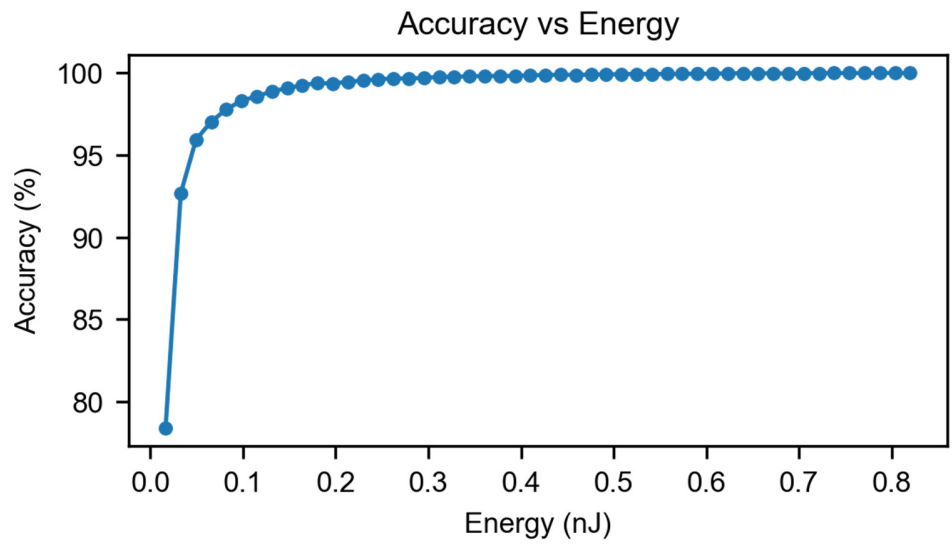

**Figure S7.** Accuracy versus cumulative energy during inference for a Hopfield network with 30% input noise.